\documentclass[a4paper,11pt]{article}

\usepackage{jheppub}
\usepackage{physics}

\usepackage{amsmath}
\usepackage{amsfonts}
\usepackage{mathrsfs}
\usepackage{amssymb}
\usepackage{amsthm}
\usepackage{mathtools}
\usepackage{bbm}

\usepackage[dvipsnames]{xcolor}
\usepackage{dsfont}

\usepackage{graphicx}
\graphicspath{ {images/} }

\usepackage[sorting=none]{biblatex}
\newcommand{\mO}{\mathcal{O}}

\newcommand{\g}{{\gamma}}
\newcommand{\eps}{{\epsilon}}

\newcommand{\lb}{ \left( }
\newcommand{\rb}{ \right) }

\newcommand{\lm}{\lambda}

\newcommand{\tN}{\tilde{N}}
\newcommand{\tlm}{\tilde{\lambda}}

\newcommand\jj[2]{\langle j^{ (#1) } j^{ (#2) } \rangle }
\newcommand\jjpm[4]{\langle j^{ (#1) }_{ #3 } j^{ (#2) }_{ #4 }\rangle }
\newcommand\jjj[3]{\langle j^{ (#1) } j^{ (#2) } j^{ (#3) } \rangle }
\newcommand\jjjpm[6]{\langle j^{ (#1) }_{ #4 } j^{ (#2) }_{ #5 } j^{ (#3) }_{ #6} \rangle }

\newcommand\V[1]{ \mathbf{V}^{#1} }
\newcommand\Vb[1]{ \bar{\mathbf{V}}^{#1} }

\newcommand\ps{{\psi}}
\newcommand\bps{{\overline \psi}}

\usepackage[T1]{fontenc}

\begin{document}

\title{A chiral limit for Chern-Simons-matter theories}

\author[a,b]{Ofer Aharony,}
\author[c]{Rohit R.~Kalloor,}
\author[a]{Trivko Kukolj}
\affiliation[a]{Department of Particle Physics and Astrophysics,
Weizmann Institute of Science, Rehovot 7610001, Israel}
\affiliation[b]{School of Natural Sciences, Institute for Advanced Study, Princeton 08540, NJ, USA}
\affiliation[c]{Department of Theoretical Physics, Tata Institute of Fundamental Research, Homi Bhabha road, Mumbai 400005, India}
\emailAdd{ofer.aharony@weizmann.ac.il, rohitrkalloor@gmail.com, trivko.kukolj@weizmann.ac.il}
\abstract{Large $N$ quasi-fermionic Chern-Simons-matter theories have an approximate higher-spin symmetry that strongly constrains their correlation functions. In particular, the 3-point functions for generic spins are combinations of 3 structures (with specific dependence on the positions and helicities), and the coupling-dependence of the coefficient of each structure is uniquely determined. In the past few years, several relations between different structures were found. In this paper we show that all the relations between the structures follow from (or, conversely, they imply) a specific form written by Skvortsov for the vertices of the dual higher-spin gravity theory on four-dimensional anti-de Sitter space, when written in spinor-helicity variables. The dual bulk theory has a specific limit where it simplifies and becomes a ``chiral higher-spin gravity theory'', and we discuss what can be said about this limit in the dual Chern-Simons-matter theories, where it involves an analytic continuation to complex couplings.}
\maketitle

\newpage

\section{Introduction}

An interesting example of the AdS/CFT correspondence is the duality between Chern-Simons-matter theories and higher-spin gravity theories on four dimensional anti-de Sitter (AdS) space. This duality was first formulated for the $O(N)$-invariant sector of the theory of $N$ free massless real scalars in $2+1$ dimensions, and for the critical $O(N)$ model related to this theory \cite{Klebanov:2002ja}. These are dual to minimal higher-spin gravity theories which only have fields of even spin, and there is a straightforward generalization to $U(N)$-invariant theories of $N$ complex scalars, dual to higher-spin gravity theories that have fields of all spins. A similar duality was then discovered \cite{Sezgin:2003pt} for $O(N)$ (or $U(N)$)-invariant theories of $N$ real (complex) free fermions, and for the critical Gross-Neveu-type fixed points associated with them. It was then discovered \cite{Aharony:2011jz,Giombi:2011kc,Aharony:2012nh} that these two dualities are the end-points of a family of dualities (which becomes continuous in the large $N$ limit), between Chern-Simons-matter theories and parity-breaking versions of higher-spin gravity. In this paper we will focus on the quasi-fermionic family of theories, whose scalar operator has dimension $\Delta_0 = 2 + O(1/N)$. This family has three different descriptions. At large $N$, these are (1) A $U(N_f)_{k_f}$ Chern-Simons (CS) theory coupled to a fermion in the fundamental representation of $U(N_f)$; (2) A $U(N_b)_{k_b}$ Chern-Simons theory coupled to the $U(N_b)$ global symmetry of the critical $U(N_b)$ model (which is the non-trivial fixed point of $N_b$ complex scalars deformed by a $|\phi^2|^2$ operator); (3) A (quantum version of a) higher-spin gravity theory \cite{Fradkin:1987ks,Vasiliev:1990en,Vasiliev:1992av,Vasiliev:1995dn,Vasiliev:1999ba} on $AdS_4$, which has a coupling $g$ (such that $n$-point vertices scale as $g^{n-2}$) and a parity-breaking-parameter $\theta$. Each of these theories also has an $O(N)$-version, which we will not discuss here. At finite $N$ all the parameters become discrete, each precise duality is slightly more complicated \cite{Aharony:2015mjs,Seiberg:2016gmd}, and each of these theories has several different versions, but that will not affect our discussion here.

In the large $N$ limit, the local operators in these theories include one operator for every integer spin (these are called ``single-trace operators''), and the products of these operators. The theories in this class have an approximate higher-spin symmetry at large $N$, meaning that the divergence of the spin $s$ currents ($s > 2$) is given by a linear combination of terms which are products of two other currents\footnote{By a slight abuse of notations, we will call all the ``single-trace operators'' currents, including the operator of spin zero.} and derivatives acting on them, with coefficients proportional to $1/N$. It was proven in \cite{Maldacena:2012sf} that the 3-point functions in theories with this property are uniquely fixed (at leading order in the large $N$ limit) up to two parameters; in the field theory descriptions these may be identified with $N$ and with the 't Hooft coupling $N/k$, and on the gravity side with $g$ and $\theta$. It is believed (but not yet proven) that all higher-point functions are also determined by the same parameters; a proof of this would imply that all three theories described in the previous paragraph are indeed equivalent at large $N$.

In \cite{Maldacena:2012sf} the 3-point functions were written as a linear combination of three different structures with a different dependence on the positions and indices of the operators, with coefficients that were different functions of the parameters (when one or more of the operators have spin zero, less structures appear). However, by writing the 3-point functions in momentum space and in spinor-helicity variables, it was noticed in a series of papers \cite{Jain:2021vrv,Gandhi:2021gwn,Jain:2021gwa,Jain:2021whr,Jain:2022ajd} that the different structures are related, and in particular that in spinor-helicity variables some of them are proportional to each other. Independently of this, the cubic vertices of the dual higher-spin gravity theories were analyzed in \cite{Skvortsov:2018uru} using light-cone coordinates and using a parameterization of the bulk fields that is closely related to the spinor-helicity basis of operators. It was suggested there that the 3-point vertices for fields with helicities $s_1, s_2, s_3$ (which can be $\pm s$ for the spin $s$ gauge fields, or $0$ for the scalar) have coefficients whose coupling dependence is $g e^{i n \theta}$ with $n = {\rm sign}(s_1) + {\rm sign}(s_2) + {\rm sign}(s_3) - {\rm sign}(s_1+s_2+s_3)$ (and they vanish when $s_1+s_2+s_3=0$). When translated into 3-point functions in the field theory, obtaining this coupling dependence requires very specific relations between the three structures described above.

In this paper we show that the relations between different 3-point functions structures take precisely the correct form to be consistent with the higher-spin gravity vertices of \cite{Skvortsov:2018uru}. This includes the relations analyzed in \cite{Jain:2021vrv,Gandhi:2021gwn,Jain:2021gwa,Jain:2021whr,Jain:2022ajd,Skvortsov:2018uru} and a few additional relations, and there are no additional relations that do not follow from this form.

It was noted in \cite{Skvortsov:2018uru} that this form of the 3-point vertices is consistent with having a specific non-unitary limit of the higher-spin gravity theories, in which only 3-point vertices where the sum of the helicities is positive are non-zero. This limit was called a ``chiral higher-spin gravity''\footnote{Such theories were constructed in flat space in \cite{Metsaev:1991mt,Metsaev:1991nb,Ponomarev:2016lrm}, and further analyzed in \cite{Skvortsov:2018jea,Skvortsov:2020gpn,Sharapov:2022wpz,Sharapov:2022nps,Skvortsov:2022syz,Didenko:2022qga,Sharapov:2022faa} (among others).}, and it was argued to have special properties such as bulk locality \cite{Ponomarev:2016lrm} and absence of bulk UV divergences \cite{Skvortsov:2020wtf}. The duality of the higher-spin gravity theories to Chern-Simons-matter theories implies that the latter should also have such a non-unitary chiral limit, in which only 3-point functions with positive total helicity are non-zero. By translating the gravity limit to the field theory, we describe the corresponding limit on the field theory side, involving a continuation of the Chern-Simons level to complex values. The limit required to obtain the chiral theories is not strictly consistent with the 't~Hooft large $N$ limit (in which the relation between the 3-point functions of higher-spin gravity and of CS-matter theories is proven), and it would be interesting to understand better the properties of this limit in the field theory.

We begin in section \ref{sec2} by reviewing the CS-matter and higher-spin gravity theories mentioned above, and what is known about their 3-point functions. In section \ref{sec:3pt} we analyze the 3-point functions for different values of the spins and helicities, and show that their different structures are related precisely in the correct way to reproduce the higher-spin gravity action in the form written in \cite{Skvortsov:2018uru}. In section \ref{sec:chiral} we analyze the ``chiral limit'' of these theories, and the subtleties with understanding it in the Chern-Simons-matter theories.

The generalization of our work to quasi-bosonic theories, where in the bulk the scalar operator obeys different boundary conditions corresponding to $\Delta=1+O(1/N)$, is straightforward. These theories have an extra coupling constant that at large $N$ only affects the 3-point function of the scalar (and higher-point functions), and which in the holographic description appears only in the boundary conditions for the scalar. The only difference between the two types of theories in the 3-point functions that do not involve 3 scalars is a Legendre transform with respect to the source of the scalar operator; in momentum space this is the same as multiplying each correlation function of the scalar at momentum $p$ by $1/|p|$. 

{\bf Note added:} During the work on this paper we became aware of \cite{Sachin}, which contains some overlapping results.

\section{A Review of quasi-fermionic Chern-Simons-Matter Theory}\label{sec2}

Quasi-fermionic (QF) Chern-Simons-Matter theories are a special case of a broad class of three dimensional CFTs involving matter coupled to Chern-Simons gauge fields. The field theories in this class are believed to have two dual descriptions, both of which can be thought of as low-energy fixed points of matter coupled to Yang-Mills-Chern-Simons gauge fields. For simplicity we describe here only the case of a single, fundamental matter field, though there is a straightforward generalization to multiple matter fields. And, we will describe the field theories in the large $N$ limit, ignoring order one shifts in the ranks or levels as well as the distinction between $U(N)$ and $SU(N)$ groups, which show up at higher orders in $1/N$. In this article, we will be working exclusively in the Euclidean signature.

As a \textit{fermionic theory}, we have a Dirac fermion $\ps$ in the fundamental representation of the gauge group $U(N_f)$, coupled to gluons $A^a_\mu$ with Chern-Simons interactions at level $k_f$:

\begin{flalign}\label{FF+CS}
 S_{FF+CS} = -\frac{ik_f}{8\pi}\int d^3 x \eps^{\mu\nu\rho} \bigg(A_\mu^a \partial_\nu A_\rho^a + \frac{1}{3}f^{abc} A^a_\mu A^b_\nu A^c_\rho\bigg) + \int d^3 x \bps\g^\mu D_\mu \ps.
\end{flalign}
In the large $N_f$ limit this theory has an 't Hooft coupling $\lambda_f\equiv N_f/k_f$ \footnote{The convention we will use in this paper is that the Chern-Simons levels are the ones appearing when the theory is regularized using dimensional regularization. They are related to the levels $\kappa_f$ appearing when flowing from a Yang-Mills-Chern-Simons theory by $k_f = {\rm sign}(\kappa_f) (|\kappa_f| + N_f)$. This implies that $|k_f| \geq N_f$ and that $|\lambda_f| \leq 1$; and similarly for the bosonic theory. \label{foot:ymcs}}.
Alternatively, one may describe the same QF theory as a \textit{bosonic theory} by starting from the $U(N_b)$ critical scalar model of $N_b$ complex scalars $\phi$ and coupling it to Chern-Simons interactions at level $k_b$:

\begin{flalign}\label{CB+CS}
 S_{CB+CS} = -\frac{ik_b}{8\pi}\int d^3 x \eps^{\mu\nu\rho} \bigg({\tilde A}_\mu^a \partial_\nu {\tilde A}_\rho^a + \frac{1}{3}f^{abc} {\tilde A}^a_\mu {\tilde A}^b_\nu {\tilde A}^c_\rho\bigg) + \int d^3 x \bigg(|{\tilde D}_\mu \phi|^2 + \sigma (\phi^\dagger\phi)\bigg).
\end{flalign}
This theory has an 't Hooft coupling $\lambda_b\equiv N_b/k_b$, and $\sigma(x)$ is an auxiliary field.

The fermionic description \eqref{FF+CS} and the bosonic description \eqref{CB+CS} are believed to be related by a strong-weak duality, and describe the same theory, with the identifications \cite{Aharony:2012nh}:
\begin{flalign}
    k_b = - N_f \ \text{sign}(k_f),  \qquad\qquad
    N_b = |k_f|.
\end{flalign}
This statement goes by the name of \textit{3D Bosonization}.\footnote{
More general versions of the duality were formulated outside the large $N$ 't Hooft limit \cite{Aharony:2015mjs,Hsin:2016blu,Aharony:2018pjn}, and there are also extensions to other gauge groups or matter content (see, for instance, \cite{Jain:2013gza,Banerjee:2013nca,Aharony:2016jvv,Cordova:2017vab,Jensen:2017bjo}), and similar dualities for supersymmetric theories \cite{Giveon:2008zn,Benini:2011mf}.}

At large $N_f$ (equivalently large $N_b$) it was shown that the QF theories with different couplings form a line of unitary conformal fixed points $0\leq \lambda_f\leq 1$ \cite{Giombi:2011kc} (by using a parity transformation we can choose $\lambda_f \geq 0$ without loss of generality). In this limit the duality relates $\lambda_b = \lambda_f - {\rm sign}(\lambda_f)$. The special endpoints of this interval are the singlet sector of the free fermion $U(N_f)$ model, reached by setting $\lambda_f \to 0$ (or equivalently $\lambda_b\to 1$), and the singlet sector of the critical $U(N_b)$ scalar model, reached by setting $\lambda_f \to 1$ (or equivalently $\lambda_b\to 0$). These two endpoints are referred to as the free fermion ($FF$) limit, and the critical boson ($CB$) limit, respectively.

The addition of Chern-Simons interactions does not introduce new local operators, but just modifies the field bilinears already present in free and critical vector models \cite{Giombi:2016zwa}. Therefore, the spectrum of ``single-trace'' operators of these theories consists of a tower\footnote{In the present case, we have a unique spin-$s$ current for each $s\geq 0$, however by choosing a different gauge group, different truncations/extensions of the spectrum are admissible \cite{Maldacena:2012sf}.} of integer spin primaries $(j^{(s\geq 0)}(x))^{\mu_1\mu_2...\mu_s}$, each transforming in the spin-$s$ representation of $SO(3)$. Here we encounter a defining feature of the QF Chern-Simons-Matter theory, which is the presence of an approximate global higher-spin symmetry \cite{Maldacena:2012sf}. While initially argued from considering higher-spin gravity duals, it has also been explicitly verified \cite{Giombi:2011kc,Giombi:2016zwa} that the higher-spin ($s > 2$) currents of the QF theories develop divergences only at subleading orders in $N$, in the form of multi-trace operators. Ignoring Lorentz indices, derivatives and numerical factors (depending on $\lambda$) that can be found, for instance, in \cite{Giombi:2016zwa}, we write this schematically as: $(\partial\cdot j^{(s\geq 3)}) \simeq \frac{1}{N_{f,b}} \sum_{s_1,s_2} j^{(s_1)}j^{(s_2)} + \mO (1/N_{f,b}^2)$, where the sum goes over $s_1+s_2 \leq s-2$.

It was shown by Maldacena and Zhiboedov in \cite{Maldacena:2011jn} that the only theories satisfying exact higher-spin symmetry are free theories (assuming a single conserved spin-2 current). Nevertheless, weakly broken higher-spin symmetry, as described above for QF theories \eqref{FF+CS}, \eqref{CB+CS} still imposes strong constraints on current correlators, with all 3-point functions belonging to a two-parameter family of correlators with a tractable amount of different kinematic structures \cite{Maldacena:2012sf}. 
These parameters can be related to the microscopic parameters in the fermionic and bosonic descriptions.
The correlators of the QF theory are controlled \cite{Maldacena:2012sf} by a large parameter $\tN$ which is of order $N_{f,b}$, and which plays the role of the effective number of degrees of freedom. The other parameter $\tlm$ controls the degree of higher-spin symmetry breaking. These parameters were identified in \cite{Aharony:2012nh,Gur-Ari:2012lgt} as:
\begin{flalign}\label{2:MZparam}
    \tlm=\tan\bigg(\frac{\pi \lambda_f}{2}\bigg)=-\cot \bigg(\frac{\pi \lambda_b}{2}\bigg); \qquad \tN=\frac{2N_f \sin (\pi\lambda_f)}{\pi\lambda_f}=\frac{2N_b \sin (\pi\lambda_b)}{\pi\lambda_b}.
\end{flalign}
With this identification, $\tlm\to 0$ corresponds to the FF limit, while $\tlm\to\infty$ corresponds to the CB limit. 
Alternatively, we can describe the correlators in the $CB+CS$ description by the bosonic $\tlm_{b}=\tan(\frac{\pi\lambda_b}{2})$ parameter, satisfying $\tlm_{b}=-\tlm^{-1}$, so that taking $\tlm_{b}\to 0$ corresponds to the CB theory.

The holographic dual of large $N$ Chern-Simons-Matter theory is a higher-spin gravity theory, whose classical equations of motion were written down in \cite{Vasiliev:1990en,Vasiliev:1992av,Vasiliev:1995dn,Vasiliev:1999ba}. There are two parity-preserving versions of these theories on $AdS_4$, and a parity-breaking parameter $\theta$ affecting 3-point and higher correlation functions that interpolates between these two versions. The 3-point vertices of these theories
were constructed in \cite{Skvortsov:2018uru}, in light-cone gauge and in a particular gauge-fixing of the bulk fields.
These interaction vertices match the dual CFT three-point functions of single-trace operators. Written in light-cone gauge, each spin-$s$ field ($s > 0$) in the bulk is represented in terms of two helicity eigenstates\footnote{We denote the bulk radial coordinate by $z$.} $\Phi_\lambda(p;z)$, where $\lambda=\pm s$ is the helicity, and $p$ is the momentum in the three dimensions of the CFT (for $s=0$ there is a single bulk field $\Phi_0(p;z)$). In the dual CFT, these fields are related to the two physical components of the corresponding spin-$s$ current $j^{(s)}_{\pm}(p)$, written in the spinor-helicity basis\footnote{Recall that symmetric traceless conserved currents in $2+1$ dimensions have only two independent components. Their components parallel to the direction of the momentum vanish by conservation, and tracelessness leaves just two independent components in the other directions. We review the spinor-helicity basis for these operators in appendix \ref{app:shbasis}.}. The final expression for the terms with up to three fields in the classical bulk action is:
\begin{flalign}\label{2:hsg}
    S &= S_2+S_3+\bar{S}_3, \\
    S_2 & = 
    \frac{1}{2} \sum_{\lm} \int dz d^3p \lb \Phi_{\lm} (p;z) \rb^{\dagger} \lb \partial_z ^2 - p^2 \rb   \Phi_{\lm} (p;z), \\
    S_3 & =g \lb e^{+ 2 \imath \theta}  \mathbf{V}^{+++} + e^{+ \imath \theta}  \mathbf{V}^{++0} + \lb  \mathbf{V}^{++-} + \mathbf{V}^{+00}\rb + e^{-\imath \theta}\mathbf{V}^{+-0} + e^{- 2 \imath \theta}  \mathbf{V}^{+--}\rb ,\label{eqn:S3c} \\
    \bar{S}_3 & =g \lb e^{-2 \imath \theta}  \bar{\mathbf{V}}^{---} + e^{-\imath \theta}  \bar{\mathbf{V}}^{--0} + \lb \bar{\mathbf{V}}^{--+}+\bar{\mathbf{V}}^{-00}\rb + e^{+ \imath \theta}\bar{\mathbf{V}}^{-+0} + e^{+ 2 \imath \theta}  \bar{\mathbf{V}}^{-++}\rb.
    \label{eqn:Sk}
\end{flalign}
The quadratic part of the action $S_2$ simply represents the action of free fields on $AdS_4$. The factors $\mathbf{V}$ and $\bar{\mathbf{V}}$ denote three-field interaction vertices explicitly given in \cite{Skvortsov:2018uru}, whose precise form will not be important for us; their indices refer to the signs of the helicities of the participating fields. The $\mathbf{V}$'s are non-zero only when the sum of the three helicities of the participating fields is positive, and the $\bar{\mathbf{V}}$'s only when it is negative.

This description of the bulk theory involves two parameters $(g, \theta)$, where $\theta \in [0, \tfrac{\pi}{2}]$ takes one from the gravity dual of the critical bosonic theory to the free fermionic one. As was pointed out in \cite{Skvortsov:2018uru}, it is natural to consider two specific limits. By dropping either the chiral vertices $S_3$ (which all have a positive sum of spinor-helicities) or the anti-chiral vertices $\bar{S}_3$ (which all have a negative sum of spinor-helicities) and setting $\theta=0$, one obtains the action of (anti-)Chiral Higher-Spin gravity on $AdS_4$.  We will discuss the interpretation of this limit in the Chern-Simons-matter theories in Section \ref{sec:chiral}.

\subsection{Two-point functions}\label{sec2:2pt}

As all the current operators are primaries of specific spin $s$ and scaling dimension $\Delta_s$, their two-point functions at separated points are fixed by conformal invariance, to a single, parity-even structure. We will normalize the currents such that the coefficients of this structure are independent of $\lambda$ and scale like $\tN$. At finite, non-zero $\tlm$, the two-point functions of spinning currents in the theories \eqref{FF+CS}, \eqref{CB+CS} develop also parity-odd contact terms. It will be useful to define the momentum space epsilon-transform of a current operator $j^{(s)}_{\mu_1...\mu_s}(p)$ \cite{Jain:2021gwa}:
\begin{flalign}\label{2:eps}
    (\eps\cdot j^{(s)})_{\mu_1...\mu_s}(p) \equiv 
    \eps_{\nu\rho(\mu_1}\frac{p^\nu}{|\vec{p}|}(j^{ (s) })^\rho_{ \mu_2 \cdots \mu_s)}(p).
\end{flalign}
The brackets in the subscript denote symmetrization with the appropriate $1/s!$ factor. The epsilon-transform appears frequently when relating different structures of two- and three-point functions of QF and QB theories. Conveniently, in the spinor-helicity basis, this operation takes the form of a helicity-dependent phase (see \eqref{A:epsSH}). The two-point functions at leading order in the large $N$ limit may now be written as:
\begin{flalign}
    \jj{0}{0}_{QF}&=\tN \jj{0}{0}_{FF}, \label{2:2pt0}\\
    \jj{s}{s}_{QF}&=\tN\bigg(\jj{s}{s}_{FF} + g_s(\tlm)\jj{s}{s}_{odd}\bigg), \text{ for }s\geq 1, \label{2:2ptS}\\
    \jj{s}{s}_{odd}&=\langle (\eps \cdot j^{ (s) }) j^{ (s) } \rangle_{FF} .\label{2:2ptCt}
\end{flalign}
Here the subscript $odd$ denotes contact terms, given explicitly by \eqref{2:2ptCt}, whose coefficient can be a general odd function $g_s(\tlm)$ of the coupling. Interestingly, perturbative computations of the vector current two-point functions in the bosonic and fermionic descriptions give different answers for these factors \cite{Aharony:2012nh,Gur-Ari:2012lgt}, specifically:
\begin{flalign}\label{2:g1fun}
    g_1(\tlm) = 
    \begin{cases}
			\tlm, & \text{in the FF+CS description }\eqref{FF+CS},\\
                \tlm_b = -\tlm^{-1}, & \text{in the CB+CS description }\eqref{CB+CS}.
    \end{cases}
\end{flalign}
So far, as far as we know, only one higher-spin contact term was computed; the stress-tensor two-point function was computed in \cite{Aharony:2012nh} in the $CB+CS$ description, to be $g_2(\tlm)=\tlm_b$. Based on this example and on the general structure of the perturbative computations leading to it, it is natural to assume that $g_s$ is independent of the spin and equal to \eqref{2:g1fun} for all $s$, and we will assume this from here on.

In general one would expect the correlation functions in dual descriptions of the same theory to agree at separate positions (and this is indeed true), but they can differ by contact terms (which can be interpreted as a change of variables between the sources in one description and in another). However, the contact terms appearing in $\eqref{2:2ptS}$ are not entirely arbitrary. When coupling the spin-$s$ currents to background gauge fields $a_s$, the contact terms \eqref{2:2ptCt} arise from background Chern-Simons terms. For the vector current $j^{(1)}$, such a term looks like $F[a]\propto ...+\frac{i\kappa_1}{4\pi}\epsilon^{\mu \nu \rho} \int d^3 x a_{1\mu}\partial_\nu a_{1\rho}$ (related to a momentum-space two-point function $\langle j^{(1)}_{\mu}(p) j^{(1)}_{\nu}(-p) \rangle \propto \epsilon_{\mu \nu \rho} p^{\rho}$), and similarly one may have gravitational Chern-Simons terms and higher-spin analogs. The quantization of the Chern-Simons levels implies that the fractional coefficients of these terms $(\kappa_s\mod 1)$ cannot be shifted by introducing local counterterms, and hence represent universal parameters of the theory \cite{Closset:2012vp}. \footnote{Note that \eqref{2:g1fun} is not enough to imply that the bulk duals of \eqref{FF+CS} and \eqref{CB+CS} differ in the value of $\theta$-terms, as the functions $\tN g_1(\tlm)$ are just the leading order contributions to $(\kappa_1\mod 1)$ in the large $\tN$ expansion. The large $N$ bosonic and fermionic descriptions have different contact terms, but their fractional parts should agree at any finite value of $N$.}

In the bulk theory, these contact-term coefficients correspond to $\theta$-angles. For an example of the $U(1)$ case, where the bulk term is proportional to $F\wedge F$, see \cite{Closset:2012vp}, and the bulk dual theory is expected to also contain $R\wedge R$ and higher-spin $\theta$-terms. The choice of prefactors\footnote{Up to a shift by integer multiples, reflecting the periodicity of $\theta$-angles.} for contact terms in \eqref{2:2ptS} then reflects different possibilities for choosing the bulk $\theta$-terms, and will play an interesting role when discussing the chiral limits of Chern-Simons-Matter theory below. 

Written in spinor-helicity variables, the $QF$ two-point functions \eqref{2:2ptS} take on a simple form for general values of spin \cite{Jain:2021gwa}. The only non-vanishing components are those with non-zero helicity:
\begin{flalign}\label{2:2ptSH}
    \langle j^{ (s) }_{ + }(q) j^{ (s) }_{ + }(-q)\rangle=
    \tN\bigg(1-i g_s(\tlm)\bigg)
    \frac{\langle \bar{1}\bar{2}\rangle^{2s}}{|\vec{q}|}; \qquad
    \langle j^{ (s) }_{ - }(q) j^{ (s) }_{ - }(-q)\rangle=
    \tN\bigg(1+i g_s(\tlm)\bigg)
    \frac{\langle 12\rangle^{2s}}{|\vec{q}|}.
\end{flalign}
Note that these correlators come with momentum-conserving delta functions, which we will choose to omit throughout this article.

\subsection{Three-point functions}\label{sec2:3pt}

It was shown in \cite{Maldacena:2012sf} that the correlation functions of the currents in QF theories may be expanded in up to three distinct structures. Assuming that the two-point functions are normalized as in \eqref{2:2pt0}-\eqref{2:2ptS}, one finds that the three-point function of three scalars vanishes (up to contact terms), while the other correlators at separated points are given by (where $s_i > 0$):
\begin{flalign}\begin{aligned}\label{2:3ptMZ}
    \jjj{s}{0}{0}_{QF} &= \tN \jjj{s}{0}{0}_{FF} 
    = \tN \jjj{s}{0}{0}_{CB},\\
    \jjj{s_1}{s_2}{0}_{QF} &= \tN\bigg(
    \frac{1}{\sqrt{1+\tlm^2}} \jjj{s_1}{s_2}{0}_{FF} 
    + \frac{\tlm}{\sqrt{1+\tlm^2}} \jjj{s_1}{s_2}{0}_{CB}\bigg),\\
     \jjj{s_1}{s_2}{s_3}_{QF} &= \tN\bigg(
     \frac{1}{1+\tlm^2} \jjj{s_1}{s_2}{s_3}_{FF} 
     + \frac{\tlm}{1+\tlm^2}\jjj{s_1}{s_2}{s_3}_{odd}\\& \qquad\qquad
     + \frac{\tlm^2}{1+\tlm^2} \jjj{s_1}{s_2}{s_3}_{CB}\bigg).
\end{aligned}\end{flalign}
The subscripts $FF$ and $CB$ denote correlators in the free fermion and critical boson theories with a single field, and correspond to the $\tlm\to 0,\infty$ limits, respectively \footnote{Note that in the large $N$ limit, three-point functions of the CB theory not containing scalar currents are identical to the free boson theory three-point functions, while the ones with scalars are related to free boson correlators by multiplying them by the scalar momentum. For a complete discussion, see \cite{Aharony:2012nh}.}. The additional $odd$ pieces do not appear in free theories, and in Section \ref{sec:3pt} we will discuss their relations to the $FF$ and $CB$ pieces.

In this article, we will write the three-point functions in the normalization common in the field theory literature, with the two-point functions scaling like $\tN$, as in \eqref{2:2pt0},\eqref{2:2ptS} (note that the precise normalization we are using for the scalar operator is different from the one used in \cite{Maldacena:2012sf}). However, in the higher-spin gravity action, the kinetic terms of the fields (giving their two-point functions at separated points) were normalized to one, so in order to compare the three-point functions to 
the bulk three-point vertices, we must use a normalization where the two-point functions scale like $\tN^0$. We will call this the \textit{gravity} normalization.

The three-point functions \eqref{2:3ptMZ} are written neglecting any contact terms, which may appear with general $\tlm$-dependences and may differ across fermionic \eqref{FF+CS} and bosonic \eqref{CB+CS} descriptions, similarly to the two-point function case. For simplicity, we will ignore the three-point contact terms in this paper.

The form of the three-point functions obtained from the bulk theory \eqref{2:hsg} is similar to \eqref{2:3ptMZ}. The vertices of the $CB$ and $FF$ theories may be obtained by setting $\theta=0, \tfrac{\pi}{2}$, respectively. In order to obtain the correct normalization for the 3-point functions we need to set $g = \tfrac{1}{ \sqrt{\tN}}$, and then one finds that the gravity action gives (in our normalization):
\begin{flalign}\begin{aligned}\label{2:3ptSk}
    \jjj{s}{0}{0}_{QF} &= \tN \jjj{s}{0}{0}_{FF} 
    = \tN \jjj{s}{0}{0}_{CB},\\
    \jjj{s_1}{s_2}{0}_{QF} &= 
    \tN \bigg(\cos(\theta) \jjj{s_1}{s_2}{0}_{CB} 
    + \sin(\theta) \jjj{s_1}{s_2}{0}_{FF}\bigg),\\
    \jjj{s_1}{s_2}{s_3}_{QF} &= \tN
     \bigg(\cos^2(\theta) \jjj{s_1}{s_2}{s_3}_{CB} 
     + \cos( \theta)\sin(\theta)\jjj{s_1}{s_2}{s_3}_{odd}\\& \qquad
     + \sin^2(\theta) \jjj{s_1}{s_2}{s_3}_{FF}\bigg).
\end{aligned}\end{flalign}
Comparing with \eqref{2:3ptMZ},
we arrive at:
\begin{flalign}\label{2:Sk2MZ}
\tan (\theta) \leftrightarrow - \tlm_b = \tlm^{-1}, \qquad \cot(\theta) \leftrightarrow \tlm.
\end{flalign}
In Section \ref{sec:3pt}, we will show that the three-point functions in the spinor-helicity basis reproduce the form of the bulk action \eqref{2:hsg}, with the above map between parameters.

\section{Consistency with the bulk action}\label{sec:3pt}

The bulk action \eqref{2:hsg} has a simple form in the light-cone gauge, which does not just imply \eqref{2:3ptSk} but also implies relations between the different structures such that each correlator with fixed helicities comes with a fixed power of $e^{i\theta}$.
In this section we discuss relations (some of which were first discussed in \cite{Jain:2021vrv,Gandhi:2021gwn,Jain:2021gwa,Jain:2021whr,Jain:2022ajd}) between various tensor-structures of the three-point functions, and show that they match precisely with what is expected from the form of the bulk action \eqref{2:hsg} (up to local contact terms). We present the relations that don't follow from these works as conjectures, and refer to explicit computations of correlation functions that support them.

We split our analysis into three cases -- with one, two, and three spinning operators, respectively.\footnote{The scalar three-point function at leading order in large $N$ is a contact term that may be set to zero.}

\subsection{\texorpdfstring{$\jjj{s}{0}{0}_{QF}$}{} }

As was discussed in Section \ref{sec2}:
\begin{align}
\jjjpm{s}{0}{0}{\pm}{}{}_{QF} 
= \tN\jjjpm{s}{0}{0}{\pm}{}{}_{FF}
= \tN\jjjpm{s}{0}{0}{\pm}{}{}_{CB}
\end{align}
for all $s$, up to contact terms. The vertices from \eqref{2:hsg} that contribute to this three-point function are:
\begin{align}
   S \sim g (\V{+00}+\Vb{-00}).
\end{align} 
We see that the three-point function in the gravity normalisation is precisely of the expected form, given the mapping between $g$ and $\tN$ 
discussed above.

\subsection{\texorpdfstring{$\jjj{s}{s'}{0}_{QF}$}{} }
We set $s \geq s'$ without loss of generality. Excluding contact terms, the correlator is determined by two pieces \cite{Maldacena:2012sf}:
\begin{align}
\jjj{s}{s'}{0}_{QF} & = \tN \lb \frac{1}{\sqrt{1 + \tilde{\lm}^2} } \jjj{s}{s'}{0}_{FF} + \frac{ \tilde{\lm}}{\sqrt{1 + \tilde{\lm}^2} }\jjj{s}{s'}{0}_{CB} \rb    .
\end{align}
It was noted in \cite{Jain:2021whr} (and supported by an explicit calculation of $\jjj{3}{1}{0}$ there) that for various helicities, the two structures satisfy:
\begin{align}
    \langle j^{(s)} _{a}j^{(s')}_{\pm} j_0 \rangle_{FF} & = 
    \pm \imath  \langle j^{(s)}_{a} j^{(s')}_{\pm} j_0 \rangle_{CB}, 
    \label{ss0Asmp}
\end{align}
independent of the helicity $a = \pm$ of $j^{(s)}$ (but note that this is not necessarily true if we exchange $s$ and $s'$).\footnote{The relation was written down in \cite{Jain:2021whr} using an epsilon-transform, and simplified to \eqref{ss0Asmp} using \eqref{A:epsSH}. The signs on the right-hand side of \eqref{ss0Asmp} may differ from \cite{Jain:2021whr} due to a mismatch in conventions (see Appendix \ref{app:shbasis} for our conventions). The invariant part of the claim is the form of the result, and that the \textit{same} sign works for \textit{every} $s,s'$. We have verified that the explicit calculation of $\jjj{1}{1}{0}$ in \cite{Gur-Ari:2012lgt} gives \eqref{ss0Asmp} in our conventions.} Using \eqref{2:Sk2MZ}, we now have:
\begin{align}\label{3.2:ss'0}
\jjjpm{s}{s'}{0}{a}{\pm}{}_{QF}  = \tN \lb \frac{\pm \imath + \tlm }{\sqrt{1 + \tilde{\lm}^2} } \jjjpm{s}{s'}{0}{a}{\pm}{}_{CB} \rb  
 = \tN e^{\pm \imath \theta } \jjjpm{s}{s'}{0}{a}{\pm}{}_{CB}.
\end{align}
The result \eqref{3.2:ss'0} is perfectly compatible with the form of the relevant bulk vertices:
\begin{align}
    S\sim g \lb e^{\imath \theta } \V{++0} + e^{-\imath \theta } \V{+-0} \rb + (cc),
\end{align}
where $(cc)$ stands for complex conjugation, and $\V{}$ is non-zero only when the sum of the helicities is positive.

Note that for \eqref{ss0Asmp} to be consistent when $s=s'$, all zero-total-helicity correlators in this class must vanish (up to contact terms):
\begin{align}
    \langle j^{(s)}_+ j^{(s)}_{-} j_0 \rangle_{FF,CB} = 0.
\end{align}
This was indeed observed in \cite{Jain:2021vrv}. We have also verified this in the $CB$ theory for $s \leq 6$. This agrees with the absence of zero-total-helicity vertices in the bulk action \eqref{2:hsg}.



\subsection{\texorpdfstring{$\jjj{s}{s'}{s''}_{QF}$}{}}
We order $s \geq s' \geq s''$ without loss of generality. The relevant correlator is:
\begin{align}
    \jjj{s}{s'}{s''}_{QF} & = \tN \lb 
    \frac{1}{1 + \tilde{\lm}^2 } \jjj{s}{s'}{s''}_{FF} 
    + \frac{\tlm }{1 + \tilde{\lm}^2 } \jjj{s}{s'}{s''}_{odd}  \right. \nonumber \\ &
    \left. \qquad \qquad + \frac{\tlm ^2}{ 1 + \tilde{\lm}^2 } \jjj{s}{s'}{s''}_{CB}\rb   .
    \label{eqn:jjj}
\end{align}
The odd pieces were dealt with in the analyses by Jain et. al. \cite{Jain:2021whr}. At least in the case where $s < s' + s''$, they found the following relation between the $odd$ and $FF-CB$ (which they called $homogeneous$) tensor structures:
\begin{align}\label{3.3:odd}
    \jjjpm{s}{s'}{s''}{a}{b}{c} _{odd}
    & =  
    \begin{cases}
        -\imath \jjjpm{s}{s'}{s''}{a}{+}{+} _{FF-CB}
        & b = c = + \\
        \imath \jjjpm{s}{s'}{s''}{a}{b}{c} _{FF-CB}
        & \text{otherwise}
    \end{cases}
\end{align}
irrespective of $a = \pm$.\footnote{Once again, the exact sign on the right-hand side of \eqref{3.3:odd} -- in our conventions -- may be confirmed by checking it against the collinear result for $\jjj{2}{1}{1}$ from \cite{Aharony:2012nh}.} For consistency with the bulk action, this relation must be true in general (the authors of \cite{Jain:2021whr} expect this to be true as well, but do not prove it). Using this in \eqref{eqn:jjj}, we get:
\begin{align}
    \jjjpm{s}{s'}{s''}{a}{b}{c}_{QF} & = 
    \begin{cases}
        \tN e^{-\imath (\tfrac{\pi}{2} - \theta)} \lb \frac{1}{\sqrt{1+ \tlm^2}} \jjjpm{s}{s'}{s''}{a}{+}{+}_{FF} 
    + \frac{\imath \tlm }{\sqrt{1+ \tlm^2}} \jjjpm{s}{s'}{s''}{a}{+}{+}_{CB}
    \rb    
        & b = c = + \\
        \tN e^{\imath (\tfrac{\pi}{2} - \theta)} \lb \frac{1}{\sqrt{1+ \tlm^2}} \jjjpm{s}{s'}{s''}{a}{b}{c}_{FF} 
    - \frac{\imath \tlm }{\sqrt{1+ \tlm^2}} \jjjpm{s}{s'}{s''}{a}{b}{c}_{CB}
    \rb 
        & \text{otherwise}
    \end{cases}
    \label{eqn:jjjsmplgen}
\end{align}
The relevant bulk vertices are:
\begin{align}
    S \sim g \lb 
    e^{2 \imath \theta } \V{+++} +
    \V{++-}
    +
    e^{-2 \imath \theta } \V{+--} \rb
    + (cc).
    \label{eqn:bulkjjj}
\end{align}
To proceed, we must deal with either side of the triangle inequality $s \lessgtr s' + s'' $ separately. 

\subsubsection{Inside the triangle: \texorpdfstring{$s < s'+s''$}{} \label{sec:in}}


In this case, since the non-conservation of the spin $s$ current only contains (at leading order in large $N$) other currents $j^{(s_1)} j^{(s_2)}$ with $s_1+s_2 < s$, all the currents in the correlator are conserved at leading order in $1/\tN$, namely\footnote{If the divergence is applied to $s'$ or $s''$, this statement holds even outside the triangle. The difference is the effective conservation of $j^{(s)}$.}
\begin{align}
    \langle \partial . j^{(s)} j^{(s')} j^{(s'')} \rangle_{QF} 
    =
    \langle j^{(s)}  \partial .j^{(s')} j^{(s'')} \rangle_{QF} =
    \langle j^{(s)} j^{(s')}  \partial .j^{(s'')} \rangle_{QF} 
    = 0.
\end{align} 
Consistency with the bulk action requires the following:
\begin{align}
    \langle j_a^{(s)} j_b^{(s')} j_c^{(s'')} \rangle _{FF}
    & = 
    \begin{cases}
    - \langle j_a^{(s)} j_b^{(s')} j_c^{(s'')} \rangle _{CB}
    & a = b = c \\
   \langle j_a^{(s)} j_b^{(s')} j_c^{(s'')} \rangle _{CB}
    & \text{otherwise}
    \end{cases}
    \label{eqn:inAsmp}
\end{align}
The second equality was observed in several cases by the Ward identity analysis of Jain et. al. (for instance, \cite{Jain:2021vrv}), and also by the explicit computations of $\jjj{2}{2}{2}_{FF}$ and $\jjj{2}{2}{2}_{CB}$ (for instance, \cite{Maldacena:2011nz}). The latter also supports the first equality. Assuming this result, \eqref{eqn:jjjsmplgen} becomes:
\begin{align}
    \jjjpm{s}{s'}{s''}{+}{b}{c}_{QF} & = 
        \begin{cases}
            \tN e^{-\imath (\tfrac{\pi}{2} - \theta)} \lb \frac{-1 + \imath \tlm }{\sqrt{1+ \tlm^2}} \jjjpm{s}{s'}{s''}{+}{+}{+}_{CB} 
            \rb
            & b = c = +
            \\
            \tN e^{\imath (\tfrac{\pi}{2} - \theta)} \lb \frac{1 - \imath \tlm }{\sqrt{1+ \tlm^2}} \jjjpm{s}{s'}{s''}{+}{b}{c}_{CB} 
            \rb    
            & \text{otherwise}
        \end{cases}
        \nonumber \\
        & = 
        \begin{cases}
            \tN e^{2 \imath \theta} \jjjpm{s}{s'}{s''}{+}{+}{+}_{CB} 
            & b = c = +
            \\
            \tN \jjjpm{s}{s'}{s''}{+}{b}{c}_{CB} 
            & \text{otherwise}
        \end{cases}
\end{align}
as expected from the bulk action \eqref{eqn:bulkjjj} (noting that in this case $s-s'-s'' < 0$, so we should  use ${\bar{\mathbf{V}}}^{+--}$ for $b=c=-$). The other cases may be obtained by complex conjugation. 

\subsubsection{Outside the triangle: \texorpdfstring{$s \geq s'+s''$}{} \label{sec:out}}

In this case, $j^{(s)}$ is not always conserved. An explicit computation of $\jjj{4}{1}{1}_{FF}$ and $\jjj{4}{1}{1}_{CB}$ (see \cite{Jain:2021whr}) points to the following relation:
\begin{align}
    \langle j_a^{(s)} j_b^{(s')} j_c^{(s'')} \rangle _{FF}
    & = 
    \begin{cases}
    - \langle j_a^{(s)} j_b^{(s')} j_c^{(s'')} \rangle _{CB}
    & b = c \\
   \langle j_a^{(s)} j_b^{(s')} j_c^{(s'')} \rangle _{CB}
    & \text{otherwise}
    \end{cases}
    \label{eqn:outAsmp}
\end{align}
which we conjecture to be true for all $s,s',s''$ outside the triangle. Our hypothesis is also supported by explicit computations of $\jjj{2}{1}{1}$ (see, for instance, \cite{Jain:2021vrv} or \cite{Aharony:2012nh,Kukolj:2024}). Assuming this, \eqref{eqn:jjjsmplgen} becomes:
\begin{align}
    \jjjpm{s}{s'}{s''}{+}{b}{c}_{QF} & = 
        \begin{cases}
            \tN e^{-\imath (\tfrac{\pi}{2} - \theta)} \lb \frac{-1 + \imath \tlm }{\sqrt{1+ \tlm^2}} \jjjpm{s}{s'}{s''}{+}{+}{+}_{CB} 
            \rb
            & b = c = +
            \\
            \tN e^{\imath (\tfrac{\pi}{2} - \theta)} \lb \frac{-1 - \imath \tlm }{\sqrt{1+ \tlm^2}} \jjjpm{s}{s'}{s''}{+}{-}{-}_{CB} 
            \rb
            & b = c = -
            \\
            \tN e^{\imath (\tfrac{\pi}{2} - \theta)} \lb \frac{1 - \imath \tlm }{\sqrt{1+ \tlm^2}} \jjjpm{s}{s'}{s''}{+}{b}{c}_{CB} 
            \rb    
            & \text{otherwise}
        \end{cases}
        \nonumber \\
        & = 
        \begin{cases}
            \tN e^{2 \imath \theta} \jjjpm{s}{s'}{s''}{+}{+}{+}_{CB} 
            & b = c = +
            \\
            \tN e^{-2 \imath \theta} \jjjpm{s}{s'}{s''}{+}{-}{-}_{CB} 
            & b = c = -
            \\
            \tN \jjjpm{s}{s'}{s''}{+}{b}{c}_{CB} 
            & \text{otherwise}
        \end{cases}
        \label{sssOutCons}
\end{align}
The other cases may be obtained by complex conjugation. Only the unbarred vertices of \eqref{eqn:bulkjjj} are relevant and the match is easy to see.

To complete the comparison of three-point functions we discuss the case where the spins saturate the triangle inequality $s=s'+s''$. Here the $(+--)$ correlators have zero total helicity, and must vanish for our conjecture (and for the matching to the gravity action) to be consistent. We have verified this in the $CB$ theory for low values of spin.

\section{A chiral limit for Chern-Simons-matter theories}
\label{sec:chiral}

It was noted in \cite{Skvortsov:2018uru} that the chiral (i.e. positive total helicity) three-point vertices of the bulk action (the ones appearing in \eqref{eqn:S3c}) were sufficient to construct a representation of the $AdS_4$ algebra, and this truncation was called a {\it chiral} higher-spin gravity, 
generalizing similar constructions in flat space \cite{Ponomarev:2016lrm,Skvortsov:2018jea,Skvortsov:2020wtf}. One way to obtain this theory is to drop by hand the terms in \eqref{eqn:Sk}, but the field theory interpretation of this is not clear. We will argue that one can construct a limit of the full higher-spin gravity theory that has the same effect. For simplicity we will discuss this limit in the dual Chern-Simons-matter theories, where it involves taking a specific (non-unitary) limit of $N$ and $k$ and a specific rescaling of the currents. The translation to the bulk, where it involves taking a limit of $(g, \theta)$ and rescaling the bulk fields, is straightforward given the relation between them described in section \ref{sec2}.

We would like to find a limit of the parameters $({\tilde N}, \tlm)$ and a rescaling of the currents that keeps only the 3-point functions related to \eqref{eqn:S3c} and not those of \eqref{eqn:Sk}. In general the rescaling could depend on the value of the spin, but it seems that the simplest option is to perform a universal rescaling of all $j_+^{(s)}$ operators and of all $j_-^{(s)}$ operators (independent of $s$). Then, by comparing the coefficients of the $(++-)$ vertices in \eqref{eqn:S3c} and \eqref{eqn:Sk}, it is clear that we must take a limit where $e^{i\theta} \to 0$. In the unitary higher-spin gravity theories $\theta$ is real, and correspondingly in the dual CFTs $\tlm$ (related to it by \eqref{2:Sk2MZ}) is real; however in order to obtain the chiral limit we must continue these parameters to complex values (in particular $\theta \to i\infty$), leading to non-unitary theories.

Let us denote $e^{i\theta} = \epsilon$ (it can have any phase), and assume that we take the $\epsilon \to 0$ limit while doing a rescaling such that
\begin{align}
    \tN &= {\tilde c} \epsilon^a, \\
    j^{(s)}_{+} &\to c' \epsilon^b j^{(s)}_{+}, \\
        j^{(s)}_{-} &\to c' \epsilon^c j^{(s)}_{-}, \\
    j_0 &\to c' \epsilon^d j_0.
\end{align}
By requiring that the correlation functions related to the six vertices in \eqref{eqn:S3c} remain finite, and recalling that in the 3-point functions their pre-factor is $\tN$ rather than $g$, we find that we need to have
\begin{align}
    a+2+3b& =0, \qquad a+1+2b+d=0, \qquad a+2b+c=0, \\
    a+b+2d&=0, \qquad a-1+b+c+d=0, \qquad a-2+b+2c=0.
\end{align}
There are only three linearly independent equations here, so we need one additional constraint, and requiring that the 2-point function of $j_0$ \eqref{2:2pt0} remains finite tells us that $a+2d=0$, so that we have
\begin{align}
    a=-2, \qquad b=0, \qquad c=2, \qquad d=1.
\end{align}
With this scaling all 3-point functions related to \eqref{eqn:S3c} have a finite $\epsilon\to 0$ limit, while all 3-point functions related to \eqref{eqn:Sk} go to zero, as desired. We can further choose the finite prefactors such that the 2-point function of $j_0$ is normalized to one, while all remaining 3-point functions are proportional to some new bulk chiral-higher-spin-gravity coupling constant $g'$; this will happen if we take ${\tilde c}=1/g'^2$, $c'=g'$. We will discuss the 2-point functions of $j_+$ and $j_-$ below.

In the Chern-Simons-matter theories, the limit we need to take on the parameters is then:
\begin{align}
    \tN & = -\frac{1}{g'^2} \frac{ 1 + \imath \tlm}{ 1 - \imath \tlm} \simeq \frac{1}{g'^2 \epsilon^2} \rightarrow \infty, \\
    \tlm & \rightarrow -\imath.
\end{align}
Translating this via \eqref{2:MZparam} to the couplings of the fermionic Chern-Simons-matter theories, we obtain
\begin{align}
    \lm_f = \frac{N_f}{k_f} & \rightarrow 1 -\imath \infty, \\
    k_f & \rightarrow \frac{ -\imath \pi }{ g'^2}.
\end{align}
Depending on the choice of phase of $g'$, at least one of $(N_f,k_f)$ must be complex, such that the CFT is no longer unitary. For instance, we can choose to keep $N_f$ real and positive. In this case $g'$ is imaginary, and we need to take the limit :
\begin{align}
    N_f \rightarrow \infty \nonumber \\
    k_f \in \imath \mathbbm{R}_{>0}
    \label{eqn:chi}
\end{align}
where large $|k_f|$ corresponds to having weak coupling in the gravity theory (small $|g'|$). The limit in the scalar variables takes a similar form. Alternatively, by choosing a different phase for $g'$, we can keep the Chern-Simons level real, and take the rank of the gauge group to be large and imaginary. Note that in various applications of Chern-Simons theory (such as the relation to the Jones polynomial) one uses a parameter $q$ which in our notation is $q = \exp(2\pi i / k_f)$, and the limit we are taking is $N_f\to \infty$, $q^{-N_f} \to 0$ with $q$ fixed (and close to $1$ if we want the dual gravity theory to be weakly coupled).
In this limit, we obtain a ``chiral Chern-Simons-matter'' theory, in which (after rescaling the operators as above) only positive-total-helicity 3-point functions are non-zero. 

Similarly, there is an antichiral limit where we keep only the negative-total-helicity vertices. This can be obtained by taking $\theta \to -\theta$, and exchanging the rescaling of $j^{(s)}_+$ with that of $j^{(s)}_-$. In the field theory this means $\lambda_f \to i\infty$, with $k_f$ having an opposite phase.

Recall that we defined $k_f$ as the Chern-Simons level when regularized by a dimensional reduction scheme; it is not so clear what $k_f$ behaves like when flowing from a Yang-Mills-Chern-Simons theory, since the relation between the two $k$'s involves ${\rm sign}(k_f)$ and it is not clear how to continue this to complex $k$'s.

\subsection{Two-point functions and contact terms}

The only remaining correlators to consider are the two-point functions of $j^{(s)}_+$ and of $j^{(s)}_-$, which are more subtle since as discussed in section \ref{sec2} they depend on the choice of contact terms.

Since in the limit we are taking ${\tilde N} \propto 1/\epsilon^2$, obtaining a finite limit for the 2-point function \eqref{2:2ptSH} of $j^{(s)}_+$ requires having $(1 - i g_s) \propto \epsilon^2$. Clearly this requires the contact term $g_s$ to become complex, but since we are already taking the physical couplings to be complex, it is not surprising that the contact term coefficients (related to the levels of Chern-Simons terms for background fields) also become complex. Indeed, it is easy to see that if we choose the contact terms to take their values \eqref{2:g1fun} computed from perturbation theory, either in the fermionic description of the Chern-Simons-matter theories, $g_s(\tlm) = \tlm$, or in the bosonic description, $g_s(\tlm) = - 1 / \tlm$, then we have $(1 - i g_s) \propto \epsilon^2$, so that with either one of these natural choices the 2-point function of $j^{(s)}_+$ remains finite.

Before our rescaling, with this choice of contact terms the two-point function of $j^{(s)}_-$ diverges, but after the rescaling we discussed above (needed to obtain finite 3-point functions) we find that this two-point function goes to $0$. Since we are in a non-unitary theory, there is nothing inconsistent about having operators with vanishing 2-point functions and finite 3-point functions, and this is what we obtain.

Note that since we are taking different rescalings for different spinor-helicity components of the fields, it is not so obvious how to define operators in position space in our theories. The two-point functions at separated points, which are proportional to ${\tilde N}$, diverge in our limit (for all spins) before we do any rescaling, and it is not clear how to interpret the rescaled operators directly in position space.

Our discussion so far is not completely self-consistent, since we considered contact terms in the two-point functions (which play an important role in making them finite in our limit), but we ignored all contact terms in the three-point functions. Indeed, not only do contact terms appear in perturabtive computations of 3-point functions (see, for instance \cite{Aharony:2012nh,Gur-Ari:2012lgt}), but the higher-spin Ward identities even relate some contact terms in two-point functions to contact terms in three-point functions \cite{Kukolj:2024}, so that we should include at least these contact terms, if we want our limit to be consistent with the approximate higher-spin symmetry. We believe that including these contact terms will not modify the consistency of the limit we described here, but we leave their full analysis to future work.

In this paper we only considered two-point and three-point functions; it would be interesting to study higher-point functions, and to see if they are well-defined in the chiral limit discussed here, and if they agree with the expectations from the chiral higher-spin gravity theory. It would also be interesting to study how correlation functions of non-local operators as in \cite{Gabai:2022vri,Gabai:2022mya,Gabai:2023lax} behave in our limit.

\subsection{Defining the limit of Chern-Simons-matter theories}

Our discussion above led us to consider a specific complex limit of Chern-Simons-matter theories. There are two major subtleties in understanding the behavior of these theories in this limit.

One issue is that $SU(N)$ Chern-Simons theories are well-defined for positive integers $N$ and for integer $k$, while our limit requires an analytic continuation to complex values (for one or both of these). Clearly, already for pure Chern-Simons theories, such a limit is not uniquely defined,
and it is not clear if there is any preferred limit. Another place where a similar continuation appears (for pure Chern-Simons theories) is in the duality of Chern-Simons theories to topological string theories \cite{Gopakumar:1998ki,Ooguri:2002gx}, where the string coupling $g_s$ is mapped (in our convention) to $i/k$, and the 't Hooft coupling $N/k$ is related to a complex parameter (related to a combination of the size of a 2-cycle and the $B$ field on it). It would be interesting to understand if the analytic continuation required in that context can be used also for the chiral higher-spin gravity theories; unfortunately it is subtle to describe the coupling of the Chern-Simons theory to matter fields in the topological string language \cite{Aganagic:2017tvx,Aharony:2019suq} so it is not clear how to check this.

The second issue is that even though the correlation functions we discussed were computed only in the 't Hooft limit of large $N$, $k$ with fixed $N/k$, the limit we eventually need to consider for the chiral higher-spin gravity is different, with large $N$ and fixed $k$. This limit is less well-understood, with no good methods to perform computations, and there is no reason why non-planar diagrams could not lead to significant corrections to the correlation functions we used, that may modify the results. From the bulk point of view the chiral higher-spin gravity theory seems to have special properties which may make it well-defined (and weakly coupled in the small $g'$ / large $k$ limit), and clearly also in the CS-matter theories such a limit (with only positive-helicity correlators) should have many simplifying properties.
The known leading non-planar corrections to correlators, such as the corrections to 2-point functions of spin $s>2$ operators related to their anomalous dimensions computed in \cite{Giombi:2016zwa}, naively diverge in the limit discussed above. However, since our limit treats different-helicity-components differently, and since contact terms play an important role in defining it, it is not clear if this naive computation of the non-planar corrections is relevant in our limit or not (recall that the 2-point function at separated points diverges also at the leading planar order, as discussed above, but this is fixed by including the contact terms and the rescaling). It is also possible that a resummation of the non-planar corrections could behave differently. It would be interesting to understand how to think about the full correlation functions in our limit, and to see if they are well-defined.

\section*{Acknowledgments}
We would like to thank Sachin Jain, Dhruva K. S., and Evgeny Skvortsov for useful discussions and for coordinating the submission of this manuscript with their upcoming paper. RRK and TK would also like to thank Erez Urbach for discussions.
This work was supported in part by Israel Science Foundation grant no. 2159/22, by Simons Foundation grant 994296 (Simons Collaboration on Confinement and QCD Strings), by the Minerva foundation with funding from the Federal German Ministry for Education and Research, by the German Research Foundation through a German-Israeli Project Cooperation (DIP) grant ``Holography and the Swampland'', and by a research grant from Martin Eisenstein. OA is the Samuel Sebba Professorial Chair of Pure and Applied Physics. The work of RRK was also supported by the Infosys Endowment for the study of the Quantum Structure of Spacetime. He would also like to acknowledge his debt to the people of India for their generous and steady support to research in the basic sciences.


\appendix

\section{The spinor-helicity basis}
\label{app:shbasis}

We briefly review the basics of the spinor-helicity formalism in three dimensional flat space. For a more comprehensive overview of spinor-helicity variables in three and four dimensions, we direct the reader to \cite{Baumann:2020dch,Jain:2021gwa,Jain:2021vrv} and referenes therein.

We start by embedding a general Euclidean three-vector $\vec{p}$ as a null four-vector $\vec{p}\to p_\mu = \big(|\vec{p}|,\vec{p}\big)$ in four dimensional Minkowski space. The corresponding Lorentz group may be decomposed as $SO(1,3)=SL(2,\mathbb{C})_{undot} \otimes SL(2,\mathbb{C})_{dot}$ with an additional reality condition imposed. The undotted and dotted subgroups are generated by\footnote{We use the notation $\sigma^\mu=(\mathds{1},\sigma^i)$, where $\sigma^i$ are Pauli matrices.}  $(\sigma^\mu)^\beta_\alpha$ and $(\sigma^\nu)^{\dot{\beta}}_{\dot{\alpha}}$, respectively. The spinor indices can be raised and lowered by contracting with the invariant epsilon tensors $\eps^{\alpha\beta}\in SL(2,\mathbb{C})_{undot}$ and $\eps^{\dot{\alpha}\dot{\beta}}\in SL(2,\mathbb{C})_{dot}$. We can now identify each four-vector as the outer product of two spinors:
\begin{flalign}
    p_\mu \to p_{\alpha\dot{\alpha}} = p_\mu (\sigma^\mu)_{\alpha\dot{\alpha}} = \lambda_\alpha \bar{\lambda}_{\dot{\alpha}}.
\end{flalign}
When restricting ourselves to three-dimensional vectors of $SO(3)$, we can introduce an orthogonal timelike vector $\tau_\mu = (1,\vec{0})$, giving us a way to identify dotted and undotted indices via $\mathds{1}_{\alpha\dot{\alpha}} = -\tau_\mu (\sigma^\mu)_{\alpha\dot{\alpha}}$. This reflects the fact that $SO(3)$ only preserves a single $SL(2,\mathbb{C})_{undot}=SL(2,\mathbb{C})$ subgroup. Thus to each spinor $\lambda_\alpha \in SL(2,\mathbb{C})$, we associate a barred spinor $\bar{\lambda}_\alpha=-\lambda_{\dot{\alpha}} \eps^{\dot{\alpha}\dot{\beta}}\mathds{1}_{\dot{\beta}\alpha}$, so that: 
\begin{flalign}
    \vec{p}=\frac{1}{2}(\vec{\sigma})^{\alpha\beta} \lambda_\alpha\bar{\lambda}_\beta.
\end{flalign}
For each three-vector $\vec{p}$, we can introduce two polarization vectors $\vec{z}^{\pm}$, satisfying $(\vec{z}^{\pm})^2=0$ and $\vec{z}^{\pm} \cross \vec{p}=\mp i |\vec{p}| \vec{z}^{\pm}$. In spinor-helicity variables, the polarization vectors take the form:
\begin{flalign}
    z^-_{\alpha\beta}=\frac{\lambda_\alpha\lambda_\beta}{2|\vec{p}|}; \qquad\qquad
    z^-_{\alpha\beta}=\frac{\bar{\lambda}_\alpha\bar{\lambda}_\beta}{2|\vec{p}|}.
\end{flalign}
When dealing with multiple vectors $\vec{p}_i$, to each we associate a pair of spinors $\lambda_i^\alpha$ and $\bar{\lambda}_i^\beta$. To contract spinors in an $SL(2,\mathbb{C})$ invariant way, we define the spinor dot-products as:
\begin{flalign}\label{A:spinordot}
    \langle i j\rangle = \eps_{\alpha\beta}\lambda_i^\alpha \lambda_j^\beta;\qquad
    \langle i \bar{j}\rangle = \eps_{\alpha\beta}\lambda_i^\alpha \bar{\lambda}_j^\beta;\qquad
    \langle \bar{i} \bar{j}\rangle = \eps_{\alpha\beta}\bar{\lambda}_i^\alpha \bar{\lambda}_j^\beta .
\end{flalign}
We may now use \eqref{A:spinordot} to translate expressions between momentum space and spinor-helicity variables. In practice, to obtain a specific helicity component of a correlator from its momentum space expression, we dot the $SO(3)$ indices\footnote{For the remainder of the section we denote the vector indices of $SO(3)$ by greek letters, disregarding the embedding in four dimensions.} of the currents in the following fashion:
\begin{flalign}
    \langle j_a^{(s_1)}...j_b^{(s_2)}\rangle = \langle \big((z_1^a)^{\mu_1..\mu_{s_1}} \cdot j^{(s_1)}_{\mu_1..\mu_{s_1}}(p_1)\big))...\big((z_2^b)^{\nu_1..\nu_{s_2}} \cdot j^{(s_2)}_{\nu_1..\nu_{s_2}}(p_2)\big) \rangle
\end{flalign}
Where $(z_i^a)^{\mu_1..\mu_{s_i}}=(z_i^a)^{\mu_1}...(z_i^a)^{\mu_{s_i}}$ are the polarization vectors corresponding to $\vec{p_i}$, and $a=\pm$ is the helicity sign. One can now use the expressions given in Appendix C.2 of \cite{Baumann:2020dch}, to rewrite the correlators as products of spinors. The expressions for two- and three-point functions simplify considerably, highlighting the elegance of spinor-helicity variables for massless fields. The epsilon-transform relation \eqref{2:eps} takes a particularly simple form:
\begin{flalign}\begin{aligned}\label{A:epsSH}
    (z^a)^{\mu_1..\mu_{s}} \cdot (\eps\cdot j^{(s)})_{\mu_1..\mu_{s}}(p) &=  
    -\eps_{\rho\nu(\mu_1}\frac{p^\nu (z^a)^{\mu_1}...(z^a)^{\mu_{s}}}{|\Vec{p}|} 
    \big(j^{(s)}\big)^\rho_{\mu_2..\mu_{s})}(p) \\&
    = -\frac{i a |\Vec{p}| (z^a)_\rho(z^a)^{\mu_2}...(z^a)^{\mu_s}}{|\Vec{p}|} \big(j^{(s)}\big)^\rho_{\mu_2..\mu_{s})}(p)= -ia j^{(s)}_a.
\end{aligned}\end{flalign}
Hence, the epsilon-transform in spinor-helicity variables acts as a phase. Note that, regardless of helicity, applying the epsilon transform two times consecutively just produces a minus sign. As a simple illustration, we present the different helicity components of the spin-1 current two-point function:
\begin{flalign}
    &\langle j^{(1)}_\mu (p_1) j^{(1)}_\nu(p_2) \rangle_{FF} = 
    \tN \frac{p_{1\mu}p_{1\nu}-g_{\mu\nu}p_1^2}{16 p_1}\\
    &\jjpm{1}{1}{a}{b}_{FF} = 
    \frac{\tN}{16p_1}\bigg((z_1^a\cdot p_1)(z_2^b\cdot p_1)-p_1^2 (z_1^a\cdot z_2^b)
    \bigg)\label{A:2pt1}\\
    &\jjpm{1}{1}{a}{b}_{odd} = (z_1^a)^\mu (z_2^b)^\nu \langle (\eps\cdot j^{(1)})_\mu j^{(1)}_\nu \rangle_{FF} = -ia \jjpm{1}{1}{a}{b}_{FF}
\end{flalign}
On shell, $\vec{p}_2=-\vec{p}_1$, and therefore $\vec{z}_2^{\pm*}=\vec{z}_1^\pm$. We use the identities found in \cite{Baumann:2020dch} and replace the scalar products in \eqref{A:2pt1} by spinor products. It is straightforward to see that mixed helicity components vanish. Putting together the QF two-point function \eqref{2:2ptS}, we find the two non-zero components: 
\begin{flalign}
    \jjpm{1}{1}{+}{+}=
    \frac{\tN}{256}\bigg(1-i g_1(\tlm)\bigg)
    \frac{\langle \bar{1}\bar{2}\rangle^2}{p_1}; \qquad
    \jjpm{1}{1}{-}{-}=
    \frac{\tN}{256}\bigg(1+i g_1(\tlm)\bigg)
    \frac{\langle 12\rangle^2}{p_1}
\end{flalign}
The result can be generalized to spin-$s$ currents, leading to \eqref{2:2ptSH}.

Finally, let's lay out our convention for the spinor-helicity currents by obtaining the sign of the two-point contact term from the results of \cite{Gur-Ari:2012lgt}. There, the momentum was chosen to be in the $x^3$ direction, and light-cone variables were used, with light-cone currents\footnote{We purposefully write light-cone indices as $m,p$ to distinguish them from the spinor-helicity indices $\mp$.} $j^{(1)}_{m,p} \equiv \tfrac{1}{\sqrt{2}} \lb j^{(1)}_1 \pm \imath j^{(1)}_2 \rb $. In these variables the two-point function of spin-$1$ currents was computed explicitly:
\begin{align}
    \langle j_m(-q) j_p (q) \rangle & \propto -\tN \vert q \vert -\imath \tN \tlm q,
\end{align}
where the second term is a contact term.
Given the direction of the momentum, we can
identify the current in spinor-helicity variables:
\begin{align}
    j_+ (q) = \begin{cases}
        j_p(-q), & q > 0 \\
        j_m(-q), & q < 0
    \end{cases}
\end{align}
This gives (say for $q<0$)
\begin{align}
    \langle j_+(q) j_+ (-q) \rangle & = \langle j_m(-q) j_p \rangle  \propto \tN \lb 1 -\imath \tlm \rb q, 
\end{align}
as in section \ref{sec2}.

\printbibliography

\end{document}